\documentclass[aip,aps,preprint,showpacs]{revtex4}
\usepackage{amssymb}

\usepackage{graphicx}

\usepackage[T1]{fontenc}

\usepackage[american]{babel}
\usepackage{siunitx}

\begin{document}

\title{Preferential Orientation Relationships in Ca$_2$MnO$_4$ Ruddlesden-Popper thin films}

\author{M.~Lacotte}
\affiliation{Laboratoire CRISMAT, CNRS UMR 6508, ENSICAEN, Universit$\acute{e}$ de Basse-Normandie, 6 Bd Mar$\acute{e}$chal Juin, F-14050 Caen Cedex 4, France.}
\author{A.~David}
\affiliation{Laboratoire CRISMAT, CNRS UMR 6508, ENSICAEN, Universit$\acute{e}$ de Basse-Normandie, 6 Bd Mar$\acute{e}$chal Juin, F-14050 Caen Cedex 4, France.}
\author{G.S.~Rohrer and P.A.~Salvador}
\affiliation{Department of Materials Science and Engineering, Carnegie Mellon University, 5000 Forbes Ave., Pittsburgh, Pennsylvania 15213.}
\author{W.~Prellier}\thanks{wilfrid.prellier@ensicaen.fr} 
\affiliation{Laboratoire CRISMAT, CNRS UMR 6508, ENSICAEN, Universit$\acute{e}$ de Basse-Normandie, 6 Bd Mar$\acute{e}$chal Juin, F-14050 Caen Cedex 4.}

\date{\today}

\begin{abstract}
A high-throughput investigation of local epitaxy (called combinatorial substrate epitaxy) was carried out on Ca$_2$MnO$_4$ Ruddlesden-Popper thin films of six thicknesses (from 20 to 400 nm), all deposited on isostructural polycrystalline Sr$_2$TiO$_4$ substrates. Electron backscatter diffraction revealed grain-over-grain local epitaxial growth for all films, resulting in a single orientation relationship ($OR$) for each substrate-film grain pair. Two preferred epitaxial $ORs$ accounted for more than 90 \% of all ORs on 300 different microcrystals, based on analyzing 50 grain pairs for each thickness. The unit cell over unit cell $OR$ ([100][001]$_{film}$ $\parallel$ [100][001]$_{substrate}$, or $OR1$) accounted for approximately 30 \% of each film. The $OR$ that accounted for 60 \% of each film ([100][001]$_{film}$ $\parallel$ [100][010]$_{substrate}$, or $OR2$) corresponds to a rotation from $OR1$ by 90$^{\circ}$ about the a-axis. $OR2$ is strongly favored for substrate orientations in the center of the stereographic triangle, and $OR1$ is observed for orientations very close to (001) or to those near the edge connecting (100) and (110). While $OR1$ should be lower in energy, the majority observation of $OR2$ implies kinetic hindrances decrease the frequency of $OR1$. Persistent grain over grain growth and the absence of variations of the $OR$ frequencies with thickness implies that the growth competition is finished within the first few \si{\nano\meter}, and local epitaxy persists thereafter during growth. 

\end{abstract}


\maketitle

\newpage

\section{Introduction}

Transition-metal oxides attract attention because of their fascinating properties, including superconductivity,\cite{Tranquada,vonZimmermann,Dwivedi,Bednorz} magnetism,\cite{Terakura,Terakura2,Greenblatt,Kakol,Copie} ferroelectricity,\cite{Lee,Nakhmanson,Scott} or insulator-to-metal transitions.\cite{Qiu,Nistor,Mahesh} There has been a growing activity in the development of thin film transition metal oxides because of their relevance to device applications, because properties can be modified owing to thin film strains, especially epitaxial coherency strains, and because metastable polymorphs / phase arrangements can be fabricated as thin films.\cite{Luders, Singh}  Nevertheless, most of the investigations of epitaxial films have used low-index commercially available single-crystal substrates for growth, which are an extremely limited region of epitaxial orientation space (though it continues to be richly mined).\cite{Locquet,Si,Bozovic,Madhavan, Ghosez} By broadening the region of epitaxial orientation space available to experimentalists, including both new substrate structures and orientations, we can develop a deeper understanding of epitaxial growth, of strain engineering anisotropic functional properties, and of phase stability for thin layers, all of which are closely linked to specific characteristics of the substrate surface. 

Towards these ends, we have been developing an approach called combinatorial substrate epitaxy (CSE), wherein a film is deposited on the polished surface of a polycrystalline ceramic substrate.\cite{CSE1,CSE2,CSE3,CSE4,CSE5,C2MO} We have demonstrated that many films grow in a locally epitaxial fashion such that each grain of the polycrystalline substrate can act as an independent single crystal substrate with a specific crystallographic orientation, resulting in there being thousands of potential substrates in any given film deposition.\cite{CSE1,CSE2,CSE3,CSE4,CSE5,C2MO} To compare the CSE approach with growth on commercially available single crystals, we previously fabricated polycrystalline sapphire Al$_2$O$_3$\cite{ALO} and perovskite LaAlO$_3$ and deposited complex oxides upon them.\cite{CSE4,CSE5} The misfit layered Ca$_3$Co$_4$O$_9$ grew in good local epitaxial registry on polycrystalline sapphire Al$_2$O$_3$ and exhibited interesting thermoelectric properties.\cite{CSE4} BiFeO$_3$ films grown on polycrystalline LaAlO$_3$ substrates exhibited high-quality grain-over-grain local epitaxial growth on all substrate grains, regardless of surface orientation.\cite{CSE5} Piezoforce microscopy was used to image and switch the piezo-domains, and the results were consistent with the relative orientation of the ferroelectric variants with the surface normal. Moreover, films on LaAlO$_3$ substrate crystals whose surface orientations were near the (100) exhibited strain dependent phases, behavior similar to films grown on analogous single crystals.\cite{CSE5}
   
Surprisingly, we have observed with CSE that only a small number of epitaxial orientation relationships ($ORs$) are required to describe the epitaxial growth over all of orientation space.\cite{CSE1, CSE3,C2MO} At first this was observed for simple oxides, such as TiO$_2$ and Fe$_2$O$_3$, grown on polycrystalline perovskite substrates.\cite{CSE1,CSE3} These observations were rationalized because they satisfied the continuation of the eutactic (nearly close packed) stacking between the two different structures, regardless of the interface plane. More recently we observed a similar effect for a complex layered oxide, namely the Ca$_2$MnO$_4$ Ruddlesden-Popper (RP) phase.\cite{C2MO} Over 95 \% of the 49 grains investigated for a 30 \si{\nano\meter} thick Ca$_2$MnO$_4$ film grown by pulsed laser deposition (PLD) on polycrystalline spark plasma sintered (SPS) isostructural Sr$_2$TiO$_4$ substrates exhibited one of only three epitaxial $ORs$.\cite{C2MO} The first $OR$ was the unit-cell over unit-cell $OR$, occurring for $\sim$ 41 \% of the grains; the second $OR$ was rotated from the first by 90 $^{\circ}$ about the a-axis, accounting for $\sim$ 35 \% of grains; and the third $OR$, was rotated from the first by $\sim$ 90 $^{\circ}$ about the 110-axis, accounting for $\sim$ 20 \% of grains. The latter two $ORs$ are intriguing since they should be higher in energy than the first, but together they occurred more frequently than the low-energy $OR$. These results imply kinetics play a role in determination of the frequency of $ORs$, which is consistent with what is known for high-quality growth of RP phases on single crystal perovskite surfaces.\cite{Gutmann,Zurbuchen,Tian,Palgrave,Yan}
 
Because kinetic challenges can manifest themselves during different stages of growth, and can be a function of orientation, a question remains as to whether the $ORs$ observed for Ca$_2$MnO$_4$ films occur during the initial nucleation stages or during continued growth, or whether they vary through a grain during growth. One of the limitations of all the initial CSE investigations is that they only dealt with films of a single thickness grown in a single condition, such as the 30 nm thick CSE film of Ca$_2$MnO$_4$.\cite{C2MO} To further demonstrate the potential of CSE in the design and growth of a wide range of complex functional oxides, it is thus important to understand how epitaxial growth proceeds with thickness and as a function of surface orientation. In the current investigation, we have grown a series of six Ca$_2$MnO$_4$ films, with different thicknesses ranging from 20 to 400 nm, by PLD on polycrystalline SPS Sr$_2$TiO$_4$ substrates, and we correlate the frequency of different $ORs$ with thickness and with substrate surface orientation for 300 microcrystalline substrates (50 each from six thicknesses). 
 
\section{Experimental}


Polycrystalline Sr$_2$TiO$_4$ samples were synthesized from commercial powders of SrCO$_3$ and TiO$_2$ (Cerac, with 99.5 \% and 99.9 \% purity, respectively). These precursors, weighed in stoichiometric proportions and homogenized by ball-milling for $\sim$ 1 hour, were calcined 1 hour at 1200 $^{\circ}$C. To reduce the grain size, these calcined powders were ground in an agate mortar before sintering by SPS. In this apparatus (Struers Tegra Force-5), powders were inserted in a cylindrical graphite die between two punches, and protected from external contamination by graphite papers. After 20 minutes at 1100 $^{\circ}$C under a uniaxial load of 50 MPa, a highly dense pellet of 20 mm diameter was obtained, whose phase purity was confirmed using x-ray diffraction. All substrates presented in this study (dimensions $\sim$ 5 x 2 x 2 mm$^3$) were extracted from the same pellet. The samples were cut along the direction of the applied pressure in SPS. Substrates are therefore assumed to present the same grain size distribution and the same density, and are also assumed to be directly comparable. Because EBSD characterization and film growth require extremely flat surfaces of high crystalline quality, each Sr$_2$TiO$_4$ substrate was subjected to meticulous polishing steps. First, several SiC papers of grain sizes down to 10 $\mu$m were used to get surfaces with no cutting or polishing marks. Second, diamond liquid pastes of grain sizes 3 $\mu$m and then 1 $\mu$m were employed to obtain mirror-like surfaces ready for film deposition or EBSD analysis.\cite{C2MO}

Ca$_2$MnO$_4$ targets used for film depositions were sintered by classical solid state routes. Thin films of six different thicknesses (20, 40, 90, 150, 300 and 400 nm) were deposited onto the surfaces of as-prepared Sr$_2$TiO$_4$ substrates. Depositions were performed at a temperature of 750 $^{\circ}$C, an O$_2$ pressure of 1.10$^{-3}$ mbar, a laser repetition rate of 2 Hz and a target-to-substrate distance of 50 mm. The deposition rate is estimated to 0.1 \AA/pulse (based on a transmission electron microscopy investigation of a 100 nm thick film). The deposition temperature was optimized, corresponding to the maximum in average image quality in EBSD of films deposited at different temperatures.\cite{C2MO} (Note a similar 30 nm film, deposited separately from the current series, was discussed in that previous publication.\cite{C2MO} It will be discussed here as a comparison film.) Film compositions were verified using energy dispersive x-ray spectroscopy (EDS). The formation of the RP phase was confirmed by grazing-incidence x-ray diffraction (GXRD).\cite{C2MO}

All substrates and films were characterized by EBSD, using the Orientation Imaging Microscopy software (OIM$^{TM}$ v.6.2 from EDAX-AMETEK, Inc.). For each substrate/film pair, EBSD analysis was performed at the same place, before and after film deposition. For a direct comparison of substrates and films, the following conditions were typically used: an SEM voltage of 20 kV, an aperture of 120 $\mu$m and a working distance of 15 mm. Inverse pole figure (IPF) maps of the surfaces of substrates and films were recorded with a beam step size of 0.3 $\mu$m using a hexagonal grid, which allows a better reconstruction of grain boundaries as compared to the square grid. To "clean" the data (remove incorrectly indexed points), points with a confidence index (CI) below 0.15 or having a misorientation angle greater than 5$^{\circ}$ compared to neighboring points were removed, and colored black. 

 \section{Results}


Fig. 1 presents [001] IPF maps of the surface of different Sr$_2$TiO$_4$ substrates (Fig. 1a, c, and e), as well as the corresponding Ca$_2$MnO$_4$ films grown upon them, with thicknesses of 40, 90, and 150 nm (respectively given in Fig. 1b, d, and f). For sake of clarity only three of the six films are presented here (a similar 30 nm film has been discussed elsewhere\cite{C2MO}). The color code is given by the unit triangle in Fig. 1g. The substrate/film pairs (namely Fig. 1a/b, c/d, e/f) are recorded exactly at the same area, before and after film deposition and the black boxes highlight such areas. Each substrate or film grain presents a nearly uniform orientation, though a minority of grains are poorly indexed and are colored black (the black region in Fig 1a/b is a scratch deliberately added as a fiducial mark). It is also evident that all film grains have grown in a grain-over-grain fashion relative to the substrate grains underneath (i.e., the shapes of grains are identical), even if some images are affected by drift in the slow scan direction (which is due to a charging effect during the scan).

When comparing different grain pairs, one can notice that some grains (examples are marked with *'s) have the same colors, i.e., the same orientation is adopted by both substrate and film.  However, some film grains (examples are marked with +'s) have completely different colors to the substrate on which it grew, i.e., the film has a completely different orientation than the substrate. Overall, the IPFs of the films are tinted blue, purple, and green, even though the substrates all have uniformly colored set of grains (there is no preferred orientation). This implies a significant fraction of the film grains have changed their orientation (color). The color of such grains indicates that they have grown with orientations along the arc between [100] and [110], and around the [110] in general. This cursory analysis indicates that the number of film grains that exhibit identical orientations to the substrate, i.e., a unit-cell-over-unit-cell growth mode, are less frequent than those that change their orientation, even though film grains grow in a grain-over-grain fashion with a single orientation per grain. Also, the colors are relatively similar between these three films of different thicknesses.


Fig. 2 depicts standard stereographic triangles showing the orientations corresponding to a large number of substrate and film grains for films of four thicknesses: (a/b) 20, (c/d) 40, (e/f) 90, and (g/h) 150 nm. Within the stereographic triangles, each point corresponds to the average orientation of a particular grain (the orientation spread within any grain is about the size of the points in this image). In general, the distributions of orientation angles within a grain were greater in the film grains than in the corresponding substrate grains, which is likely due to the relaxation of coherency strains. One can observe that for substrates, Fig. 2a, c, d, and e, there is a random distribution of points throughout the stereographic triangles. This confirms that all substrates are uniform and do not present a particular texture, as expected from a normal ceramic and consistent with the images shown in Fig. 1. 

The situation is different for the films, where the points in the central region of the triangle occur less frequently than in the substrate, and the number near the [110], and in the band between the [110] and [100], occur more frequently than in the substrate. There is a small cluster of points in the films near the [001] orientation as well. These observations are consistent with IPF maps of Fig. 1, again indicating that the orientation of film grains differ from those in the substrate, favoring (disfavoring) blue, purple, and green oriented grains (other orientations), with red grains having a significant frequency as well.


Using the methods and software of Zhang et al.,\cite{CSE1} the $ORs$ were determined for 50 substrate-film grains for each of the six thicknesses. For all 300 substrate-film grain pairs analyzed, two primary $ORs$ were identified (these were more than 90 \% of all grains on all films). These $ORs$ were described as $OR1$ and $OR2$ in our previous work,\cite{C2MO} and will be called the same herein. The orientation of the substrate (film) grains are plotted in the standard stereographic triangles in Fig. 3 as filled (open) symbols, with orientations that supported $OR1$ ($OR2$) given as circles (squares). (Each point again corresponds to the average orientation of the grain analyzed). The values plotted in Fig. 3 correspond to those obtained from films of thicknesses (a) 20, (b) 40, (c) 90, and (d) 150 nm.  

$OR1$ can be written as [100][001]$_{film}$ $\parallel$ [100][001]$_{substrate}$, describing a unit-cell over unit-cell epitaxial growth (with a small angular variation associated with strain relaxation\cite{CSE1, CSE3, C2MO}). For $OR1$, the symbols corresponding to the same substrate-film grain are close to each other (i.e., nearby closed and open red circles) in Fig. 3. These observations correspond to grains that exhibit the same colors between the substrate and film in Fig. 1. From purely and interfacial energy perspective, this $OR$ should be favored as it minimizes excess energy for the film/substrate interface. However, there are generally fewer red points than blue points in all films of Fig. 3. 

$OR2$ can be written as [100][001]$_{film}$ $\parallel$ [100][010]$_{substrate}$, describing substrate-film grain pairs that are exactly misoriented by an angle of 90 degrees about the a-axis. For $OR2$, the symbols (blue squares) corresponding to the same substrate-film grain are far away from each other and not easily correlated in Fig. 3. $OR2$ corresponds to the vast majority of grains that change color from the substrate to the film in Fig. 1, and for the emptying of orientations from the center of the triangle in Fig. 2. From purely an interfacial energy perspective, this $OR$ should be higher in energy, as it aligns the longer anisotropic c-axis of Ca$_2$MnO$_4$ (a = 3.668 \AA,  c =12.050 \AA)\cite{RP,Leonowicz,Autret} with the short b-axis of Sr$_2$TiO$_4$ (a = 3.884 \AA, c = 12.600 \AA),\cite{RP}, obviously leading to some increase in interfacial energy.  However, there are generally more blue points than red points in all films of Fig. 3.


It should be noted that outliers were observed, but were left out of the plots in Fig. 3, for clarity, and their orientations were not quantified in general (though in the prior 30 nm film a much higher percentage of outliers were observed and were part of a single third $OR$\cite{C2MO}). However, we did keep track of their observation and they were included as a single group of other orientations when determining the fractional population of $ORs$ for each film. The percentage of each of these three $OR$ groups was calculated for each film and are plotted as a function of thickness in Fig. 4. (Additionally, the values obtained from the prior 30 nm film are included\cite{C2MO}). Red circles correspond to $OR1$, blue squares to $OR2$, and the collection of other $ORs$ are represented by black triangles. It may be noted that the percentage of each $OR$ is not a significant function of thickness. For the current series of films, $OR2$ accounts for about 60 to 70 \% of grains, $OR1$ for about 25 to 35 \%, and others for 0 to 10 \% of grains. These observations reflect the data presented in Figures 1-3. Interestingly, the current series of films differ quantitatively from the 30 nm film presented previously, which had $OR2$ accounting for about 35 \% of grains, $OR1$ for about 45 \%, and others for 20 \% of grains. 
 
\section{Discussion}

This high-throughput investigation (CSE) of epitaxial growth of the complex Ca$_2$MnO$_4$ Ruddlesden-Popper (RP) phase on polycrystalline substrates of isostructural Sr$_2$TiO$_4$ reinforces some of the initial observations made using CSE. First, epitaxial grain-over-grain film growth is observed over all of epitaxial orientation space (i.e., for all substrate orientations). Second, a single orientation relationship ($OR$) exists for almost all film/substrate pairs, again regardless of the orientation of the substrate. Third, only a small number of epitaxial $ORs$ are observed, even though the RP crystal structure is relatively complex and anisotropic. Specifically, more than 90 \% of 300 quantified observations belong to only two $ORs$: a unit-cell over unit-cell $OR$ ($OR1$) that accounts for about 30 \% of the population, and an $OR2$ that is rotated from $OR1$ by 90 $^{\circ}$ about the a-axis and that accounts for about 60 \% of the population. 

This is the first CSE study that focuses on quantifying these $ORs$ as a function of thickness. Interestingly, the population of these two $ORs$ is not significantly affected by thickness, from 20 to 400 nm. This indicates that the population of $ORs$ is essentially determined before films are 20 nm thick, and that the orientations are stable during continued growth. The simplest interpretation of this observation is that the $ORs$ are determined during the initial nucleation of films on the substrate surface. The orientation dependence of the $ORs$, specifically the absence of $OR1$ in the center of the triangle, also indicates that thermodynamic and kinetic factors control nucleation off of the high index substrate surfaces, as discussed below. The important point to be reinforced is that the polycrystalline surfaces used as substrates in CSE can be treated as independent surfaces of microcrystals that result in local epitaxial growth of single orientations, and that the growth competition is dictated by the substrate surface, similar to what is known for growth on commercial single crystals.  

For all of these films, when a substrate grain has an orientation in the arc between [100] and [110] (in the color range between green and blue), the film grain that grows upon it generally maintains the same orientation: i.e., $OR1$ is observed. This can be seen in Fig. 3 by the preponderance of closed red circles in this region of the stereographic triangle, and the paucity of closed blue squares. What is interesting in this observation, is that all of these orientations are close to having the c-axis in the plane of the substrate. Such orientations would require less out-of-plane diffusion during growth to establish an electro-neutral, stoichiometric growth unit. This is well known to have significant kinetic advantages.\cite{Gutmann,Zurbuchen,Tian,Palgrave,Yan}
In fact, for $OR2$ to form in this band of orientations, the c-axis would have to generally rotate towards the substrate normal, resulting in a kinetic challenge to growth. The combination of a low-energy interface and a kinetic advantage seem to support the favoring of $OR1$ in this region of epitaxial orientation space. 

Somewhat counterintuitive to this argument is the observation that $OR1$ is found often for grains close to [001]. In this region of orientation space, the film orientation is split between $OR1$ and $OR2$. For $OR2$ film grains in this region, the ninety degree rotation would result in orientations clustered in the arc between [100] and [110]. While $OR2$ would have a kinetic advantage for growth, there would be a significant interfacial energy disadvantage. Still, the fact that films oriented near [001] can grow with $OR1$, indicates that the growth conditions do allow for a significant amount of out-of-plane diffusion required to access purely [001] oriented films. This orientation is the most kinetically challenged with respect to $OR1$ growth, but it does occur with a higher probability than most other substrate orientations. In other words, out-of-plane diffusion cannot be the only factor that favors $OR2$ in the center of the triangle. It should be noted here that the (001) plane is the lowest energy surface plane for the RP system.\cite{Zschornak} If we consider that the film surface energy is important during nucleation of films with anisotropic structures, then we can rationalize the observation of $OR1$ near the [001] orientations as arising from the combination of a low film/substrate interface, a low film surface energy, but a challenging kinetic condition. Returning to the observation of $OR1$ occurring typically for orientations between [100] and [110], this indicates that whatever the surface energy preference for (001), which might favor $OR2$ in this space, it is not enough to overcome the interfacial and kinetic preferences for $OR1$.

For all of these films, when a substrate grain has an orientation in the center of the stereographic triangle (away from arc between [100] and [110], and away from [001]), the film grain that grows upon it generally adopts $OR2$; such grains have a single $OR$ with respect to the substrate grains, represented by the 90 degrees about the a-axis of $OR1$. This is reflected in Fig. 3 by the large number of open blue squares in this region (and the low frequency of closed red circles), to an increased observation of blue and green colors in Fig. 1, and the disappearance of film orientations in the center of the triangle in Fig. 2. This can be rationalized again by considering the thermodynamic and kinetic factors involved. For $OR2$, there is an interfacial energy penalty but a kinetic preference to grow with an orientation away from [001]. Furthermore, these orientations are likely to have the highest surface energy of all, and can potentially lower their total energy by rotating to the lower index planes of $OR2$. Even if $OR1$ and $OR2$ have similar surface energies in this region, kinetic preference and low stacking fault energies will favor $OR2$. What is unknown in all of these discussions 
 is the effect of local surface roughness on lateral diffusion during growth. It is possible that high-energy surfaces in the middle of the triangle had higher kinetic barriers to lateral diffusion, favoring kinetically preferred $ORs$ to thermodynamically preferred $ORs$. 

The primary point of these discussions is that the same rationalizations used to explain epitaxial growth on single crystals can also be used on the general surfaces of high-quality polycrystals. This further reinforces that CSE growth greatly expands our understanding of epitaxial growth in a high-throughput fashion. More than 300 observations of growth were discussed in this paper, and they are internally consistent using at least 6 different substrates and depositions. However, it is evident that the 30 nm film discussed in the prior work represents a different growth condition, even though they were nominally identical (that film was grown separately from the current series). Whether the surface of substrates, details of ablation or deposition were different is hard to determine in retrospect. The different relative ratios of $ORs$ for that 30 nm film indicates that the growth conditions can indeed modify the relative ratios of $ORs$, which is of course well known in growth on single crystals. The collected set of observations indicate that growth in CSE is highly reproducible (the current series), but that the relative ratio of competitive $ORs$ can be influenced to favor one over the other. In other words, we believe that growth conditions will exist where $OR1$ (the thermodynamically preferred film/substrate interface) will be obtained with 100 \% frequency, while other conditions will exist where $OR1$ will be only found in the arc between [100] and [110]).

\section{Conclusions}

This CSE investigation of Ca$_2$MnO$_4$ growth for films of six thicknesses (from 20 to 400 nm) deposited on polycrystalline Sr$_2$TiO$_4$ revealed grain-over-grain local epitaxial growth for all films, resulting in a single orientation relationship ($OR$) for each substrate-film grain pair. Two preferred epitaxial $ORs$ accounted for more than 90 \% of all ORs on 300 different microcrystals, with the unit cell over unit cell $OR$ ([100][001]$_{film}$ $\parallel$ [100][001]$_{substrate}$, or $OR1$) accounting for approximately 30 \% of each film. The $OR$ that accounted for 60 \% of each film ([100][001]$_{film}$ $\parallel$ [100][010]$_{substrate}$, or $OR2$) corresponds to a rotation from $OR1$ by 90$^{\circ}$ about the a-axis. $OR2$ is strongly favored for substrate orientations in the center of the stereographic triangle, and $OR1$ is observed for orientations very close to (001) or to those near the edge connecting (100) and (110). The relative frequency and preferred orientations for the two $ORs$ can be rationalized by considering the thermodynamic (interfacial and surface energies) and kinetic preferences (relative amount of out-of-plane diffusion) required to obtain a specific $OR$ on a given substrate surface. Persistent grain over grain growth and the absence of variations of the $OR$ frequencies with thickness implies that the growth competition is finished within the first few \si{\nano\meter}, and local epitaxy persists thereafter during growth. 


\section{Acknowledgements}

We thank L. Gouleuf and J. Lecourt for technical support. M. Lacotte received her PhD scholarship from the Minist$\grave{e}$re de l'Enseignement Sup$\acute{e}$rieur et de la Recherche. Partial support of the French Agence Nationale de la Recherche (ANR), through the program Investissements d'Avenir (ANR-10-LABX-09-01), LabEx EMC3, and the LAFICS are also acknowledged. PAS and GSR acknowledge the support of National Science Foundation grant DMR 1206656. 

\newpage
Figure Captions 

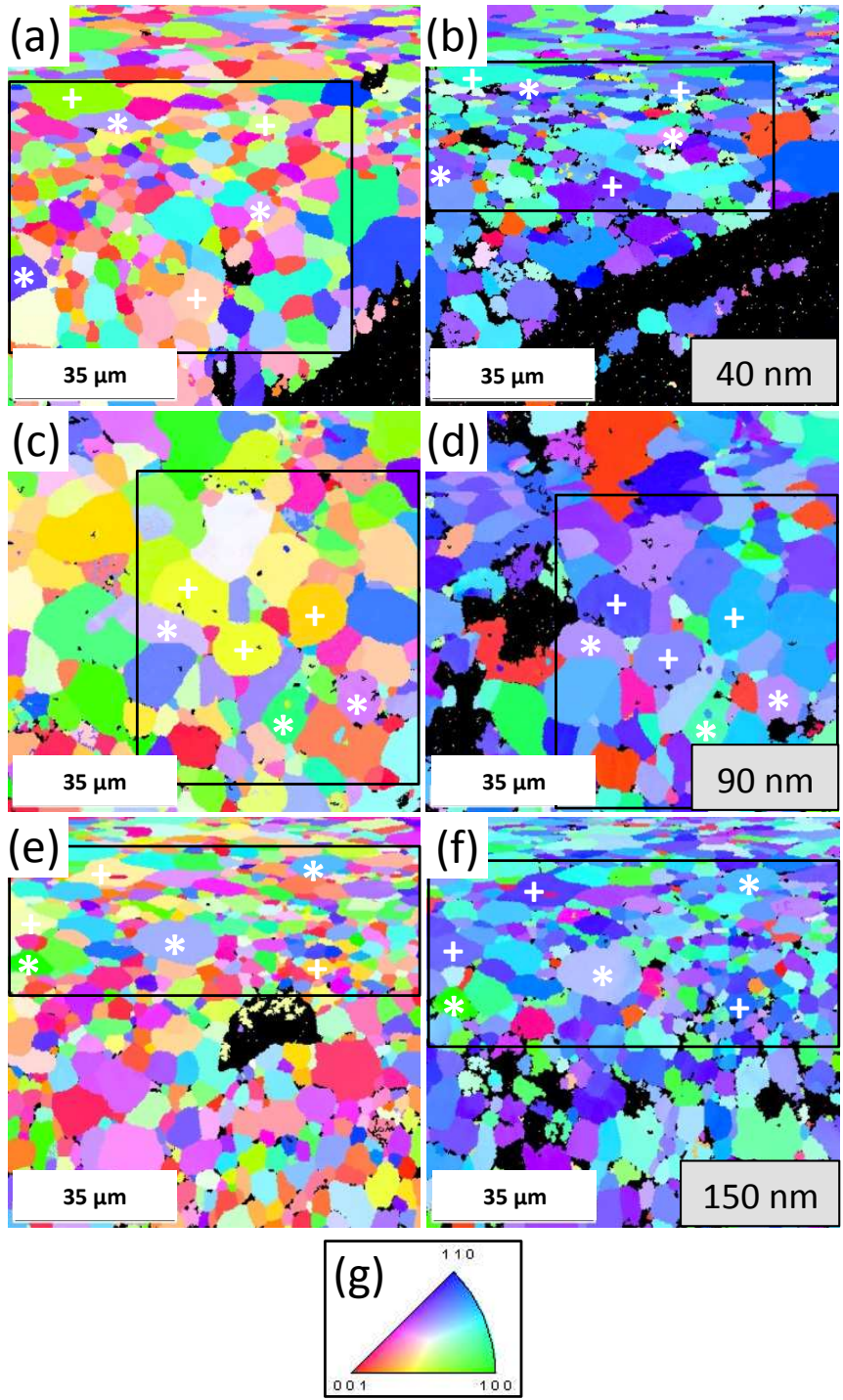
Figure 1: Pairs of substrate-film IPF maps (a/b, c/d, e/f) recorded from the same areas on Sr$_2$TiO$_4$ substrates (a,c,e) and Ca$_2$MnO$_4$ thin films of different thicknesses: (b) 40 nm, (d) 90 nm, and (f) and 150 nm. In the boxed areas, grains marked with * (+) are those who grow with OR1 (OR2), see text, as noted by the similar (different) colors between the film and substrate grains. 

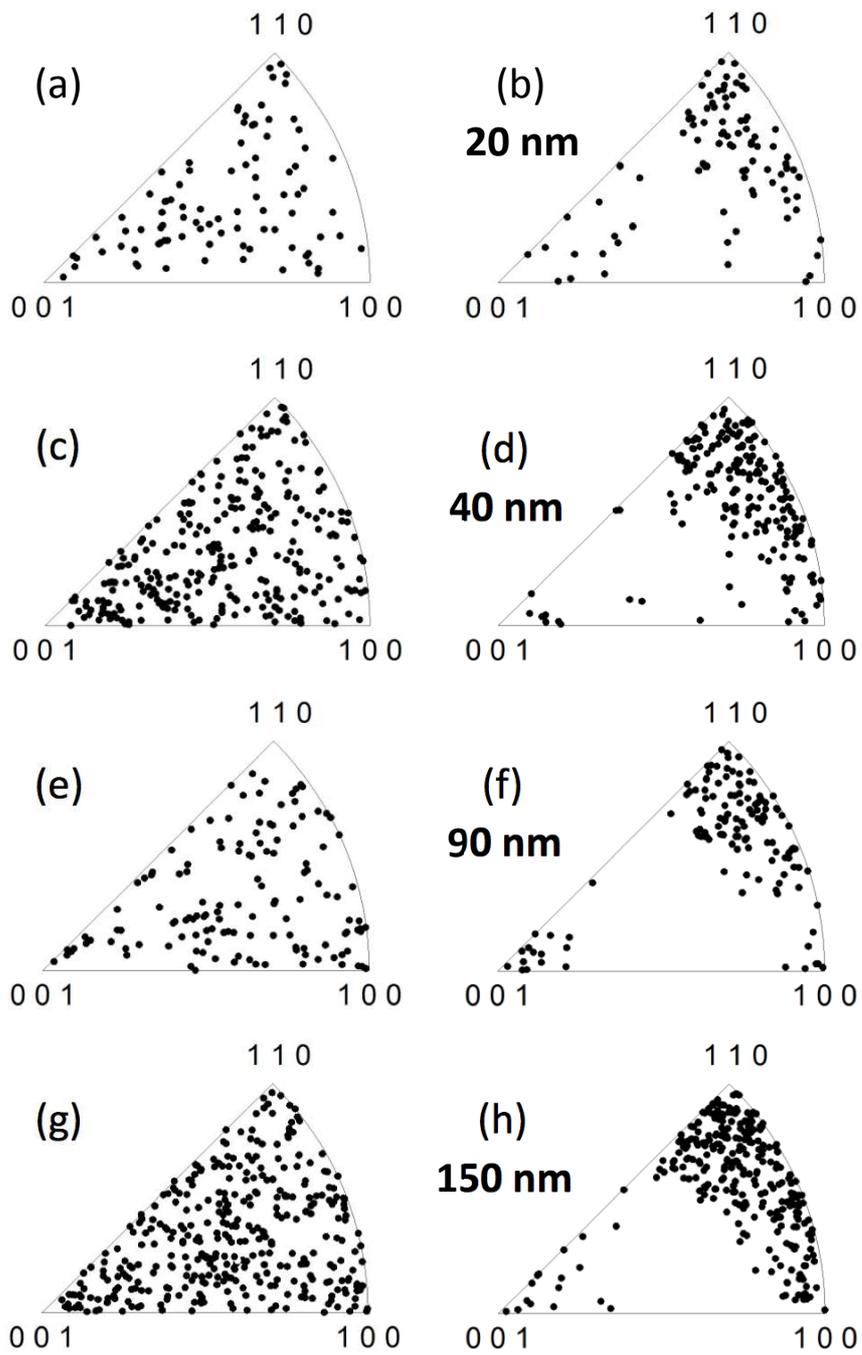
Figure 2: IPF stereographic triangles showing the orientations corresponding to the pairs of substrate-film IPF maps, with (a), (c), (e), and (g) for Sr$_2$TiO$_4$ substrates, and (b), (d), (f), and (h) for related Ca$_2$MnO$_4$ films of 20, 40, 90, and 150 nm, respectively.

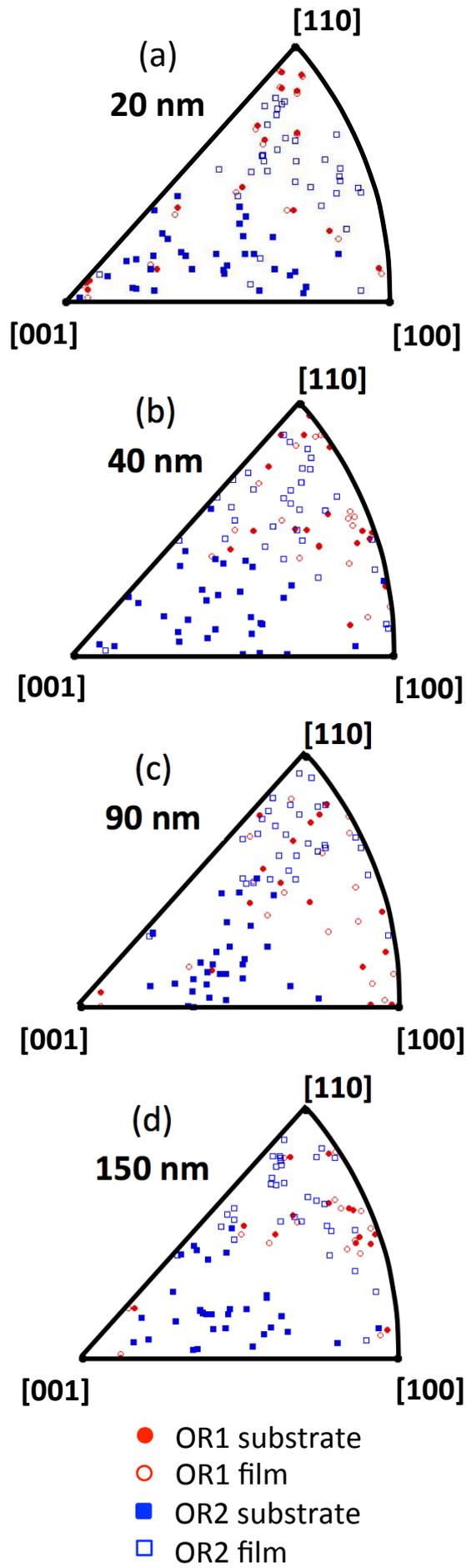
Figure 3: Standard stereographic triangles showing the $ORs$ of 50 pairs of substrate-film grains, for film thicknesses of (a) 20, (b) 40, (c) 90, and (d) 150 nm. Filled symbols are related to Sr$_2$TiO$_4$ substrate grains and open symbols to Ca$_2$MnO$_4$ film grains. Red and blue symbols correspond to $OR1$ and $OR2$, respectively.

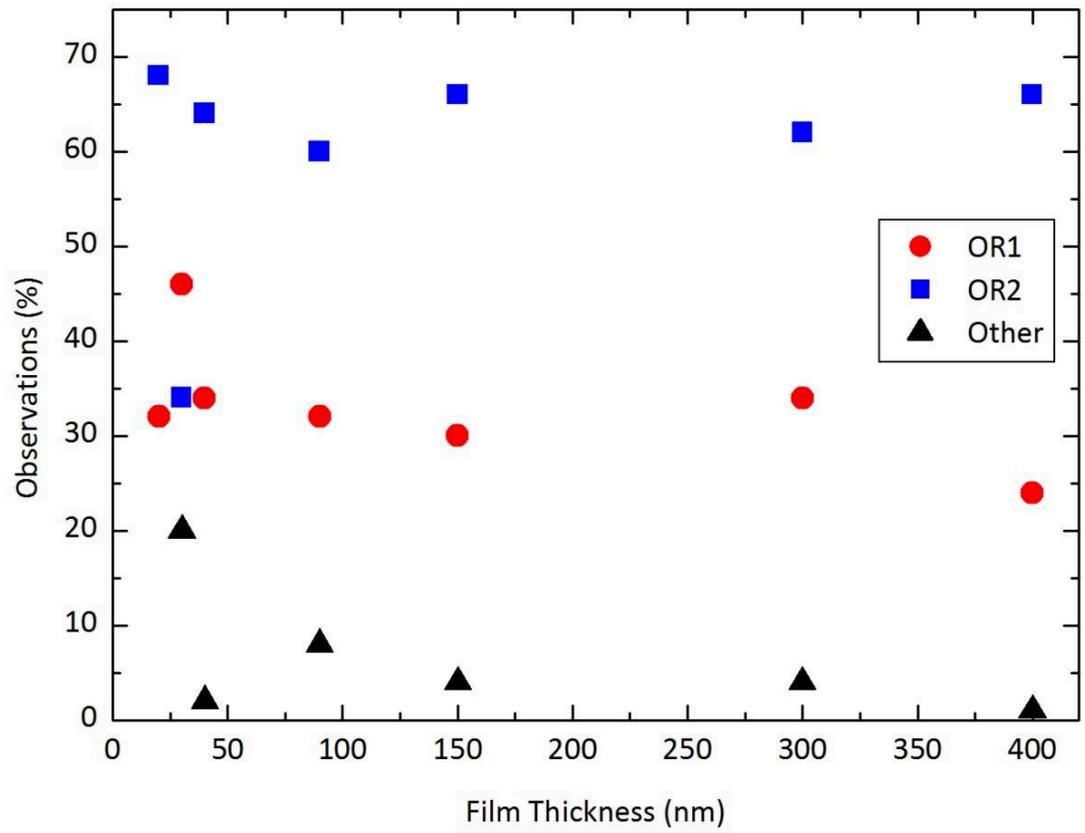
Figure 4: Fractional population of each $OR$ plotted versus the film thickness. Red circles correspond to $OR1$ and blue squares to $OR2$. All the other $ORs$ are represented by black triangles. Top and bottom insets provide the approximate relationship between the unit cells for $ORs$ 1 and 2, respectively.

\end{document}